\documentclass[12pt,a4paper,english,superscriptaddress,aps,nofootinbib]{revtex4}
\usepackage[utf8]{inputenc}
\usepackage[T1]{fontenc}
\usepackage{amsmath,amssymb,graphicx}
\makeatletter
\usepackage{babel}
\usepackage[active]{srcltx}
\usepackage{graphicx,color}
\usepackage{changebar}
\usepackage{hyperref}
\hypersetup{
	colorlinks=true,
	urlcolor= blue,
	citecolor=blue,
	linkcolor= blue,
	bookmarks=true,
	bookmarksopen=false,}

\usepackage[T1]{fontenc}
\usepackage{ae}
\usepackage{amsmath}
\usepackage{braket}
\usepackage{mathtools}
\usepackage{slashed}
\usepackage{empheq}
\usepackage{tikz}
\usepackage{multirow}
\usetikzlibrary{decorations.pathmorphing}
\usepackage[caption = false]{subfig}

\usepackage{graphicx}
\usepackage{amsfonts}
\usepackage{amsmath}
\newcommand{\be}{\begin{equation}}
	\newcommand{\ee}{\end{equation}}
\newcommand{\bea}{\begin{eqnarray}}
	\newcommand{\eea}{\end{eqnarray}}

\begin{document}

\title{String-like brane splitting in the context of $f(T,\mathcal{B})$ gravity}

\author{A. R. P. Moreira} 
\email{E-mail: allan.moreira@fisica.ufc.br}

\author{C. A. S. Almeida}
\email{E-mail: carlos@fisica.ufc.br}

\affiliation{Universidade Federal do Cear\'a (UFC), Departamento de F\'isica,\\ Campus do Pici, Fortaleza - CE, C.P. 6030, 60455-760 - Brazil}

\begin{abstract}

\vspace{0.5cm}

\noindent \textbf{Abstract:} {In this work, the influence of the boundary term $\mathcal{B}$ is analyzed in a string-like thick brane scenario in the gravity context $f(T,\mathcal{B})$.
For that, three models of $f(T,\mathcal{B})$ are proposed, i. e., $f_1(T,\mathcal{B})=T+k\mathcal{B}^{n}$, $f_2(T,\mathcal{B})=T+k(-T+\mathcal{ B})^{n}$ and $f_3(T,\mathcal{B})=T+k_1T^2+k_2\mathcal{B}^2$, where $n$, $k$ and $k_{1 ,2}$ are parameters that control the deviation from the usual teleparallelism. The first relevant result obtained was the appearance of a super-located tower in the core for energy density. Furthermore, the greater the influence of the boundary term, the new maximums and minimums appear in the energy density. All this indicates the emergence of capable structures from split to the brane. The second relevant result was obtained by analyzing the gravitational perturbations, where the effective potential presents the supersymmetric form of quantum mechanics, leading to well-localized massless modes.}


\end{abstract}

\maketitle

\section{Introduction}

The emergence of general relativity (GR) was a milestone in the evolution of modern physics. However, the GR presented some unsolved physical problems, such as the gauge hierarchy problem, the cosmological constant problem, and the problem of defining dark matter and dark energy. These physical problems were a strong motivator for the study of extradimensional models. The model proposed by Randall and Sundrum (RS) was able to assume solutions for such physical problems \cite{rs,rs2,cosmologicalconstant,darkmatter}. This model is based on the idea that our world (brane) is submerged in a five-dimensional space (bulk), where the brane can be represented by domain walls \cite{Rubakov,Visser,Gremm1999}. Thus, the study of thick brane has gained a lot of attention in recent 
 years \cite{Charmousis2001, Arias2002, Barcelo2003, Bazeia2004, CastilloFelisola2004, Navarro2004, BarbosaCendejas2005, Bazeia2007, Koerber2008,deSouzaDutra2008,Bazeia2008,conifold,Almeida2009,Dzhunushaliev2009, Cruz2013, Liu2011, Dutra2014}.

Later, inspired by the RS model, Gherghetta and Shaposhnikov 
(GS) \cite{Gherghetta}, proposed an extradimensional model where the bulk has six dimensions with axial symmetry, where the geometry resembles that of the cosmic strings \cite{Israel,Vilenkin1,Hiscock}. This model became known as string-type braneworld \cite{Navarro203,Navarro205,Kanno204,Kofinas205,Kofinas206}, and has advantages over the RS model, such as, for example, the massless Kaluza-Klein (KK) modes of the gauge field are localized on the brane without the need for any other coupling \cite{Oda,Giovannini:2001hh}, and the fermionic fields are trapped in the brane through a simple minimum coupling of gauge \cite{Liu}.
Also, the correction for the Newtonian potential due to the KK modes in the GS model is smaller than in the RS model \cite{Gherghetta}.

Currently, what has been drawing more attention is the study of these extradimensional models in modified teleparallel gravity \cite{Yang2012, Capozziello, Menezes, tensorperturbations, ftnoncanonicalscalar, ftborninfeld, ftmimetic}.
In the five-dimensional model, it was It is possible to obtain an internal structure in the thick brane scenario with only a real scalar field as a source, in addition to generating significant modifications in the location of gravity and fermions \cite{Yang2017,Moreira:2021xfe,Moreira20211,Moreira:2021wkj}. In a six-dimensional model, little is known about the influence of the modification of gravity in the brane scenario. Moreira et al, showed that the modification of the gravity $f(T)$ can cause the split in the string-like brane, generating a ring-like structure, besides causing modifications in the gravitational perturbations \cite{Moreira:2020rni,Moreira:2021fva}.

The interesting thing about modified teleparallel gravity is that the curvature has no influence on gravity, and the relevant connection is the Weitzenb\"{o}ck connection where the dynamic variable is the vielbein \cite{Hayashi1979, deAndrade1997, deAndrade1999, Aldrovandi}. In addition to torsion $T$, the boundary term $\mathcal{B}$ can influence gravity, which gave rise to the gravity model $f(T,\mathcal{B})$ \cite{Bahamonde2015, Wright2016, Bahamonde2016}. The study of this influence has attracted a lot of attention in recent years for providing a good approximation with observational data for the accelerated expansion of the universe \cite{Franco2020,EscamillaRivera2019}, in addition to providing relevant results in the study of the nature of dark energy,
cosmological perturbations and gravitational waves \cite{Bahamonde2016a, Caruana2020,Pourbagher2020,Bahamonde2020a,Azhar2020,Bhattacharjee2020,Abedi2017}.

All these results inspire us to study the influence of the boundary term and the torsion variation in a string-like thick brane scenario. The paper is organized as follows: In section (\ref{sec1}) we build the string-like thick brane scenario in the context of modified teleparallel gravity $f(T,\mathcal{B})$. Furthermore, we present the influence of the boundary term on the energy density and the respective angular and radial pressures. In section (\ref{sec2}) the influence of the boundary term on gravitational perturbations is analyzed. Finally, in the section (\ref{finalremarks}) conclusions and additional comments are presented.

\section{Brane-string-$f(T,\mathcal{B})$}
\label{sec1}
In this section, we introduce the main definitions involved in the construction of $f(T,\mathcal{B})$ gravity, in addition to presenting the modified gravitational equations for the string-like thick brane scenario.

The $f(T,\mathcal{B})$ gravity is a modification of the teleparallel equivalent of general relativity (TEGR), where gravity is generated by torsion and the dynamic variable is the well-known vielbeins $h^a\ _M$, which relates to the metric in the form 
\begin{eqnarray}
g_{MN}=\eta_{ab}h^a\ _M h^b\ _N.
\end{eqnarray}
Here the capital index $M={0,...,D-1}$ represents the coordinates of the bulk and the lower case index $a={0,...,D-1}$ represents the coordinates of the tangent space. 

Furthermore, in TEGR the relevant connection is the Weitzenb\"{o}ck connection $\widetilde{\Gamma}^P\ _{NM}=h_a\ ^P\partial_M h^a\ _N$, which relates to the Levi-Civita connection in the form \cite{Aldrovandi}
\begin{eqnarray}
\widetilde{\Gamma}^P\ _{NM}=K^P\ _{NM}+\Gamma^P\ _{NM}, 
\end{eqnarray} 
where 
\begin{eqnarray}
K^P\ _{NM}=\frac{1}{2}\Big( T_N\ ^P\ _M +T_M\ ^P\ _N - T^P\ _{NM}\Big)
\end{eqnarray}
is the contorsion tensor, being 
\begin{eqnarray}
T^{P}\  _{MN}= \widetilde{\Gamma}^P\ _{NM}-\widetilde{\Gamma}^P\ _{MN},
\end{eqnarray}
is the torsion tensor.

The TEGR Lagragian is defined as
\begin{eqnarray}
\mathcal{L}= -\frac{hT}{4\kappa_g},
\end{eqnarray} 
where $h=\sqrt{-g}$, $g$ is the determinant of the metric, $\kappa_g$ is the gravitational constant, and $T\equiv T_{PMN}S^ {PMN}$ is the torsion scalar, where \cite{Aldrovandi}
\begin{eqnarray}
S_{P}\ ^{MN}=\frac{1}{2}\Big( K^{MN}\ _{P}-\delta^N_P T^{QM}\ _Q+\delta^M_P T^{QN}\ _Q\Big),
\end{eqnarray}
is the dual torsion tensor.

It is important to note that the curvature scalar $R$ is related to the torsion scalar $T$ in the form \cite{Aldrovandi}
\begin{eqnarray}
R=-T-2\nabla^{M}T^{N}\ _{MN},
\end{eqnarray}
where
\begin{eqnarray}
B\equiv -2\nabla^{M}T^{N}\ _{MN}=\frac{2}{h}\partial_M(h T^M),
\end{eqnarray}
is defined as the boundary term.

Being a generalization of TEGR, the gravitational Lagrangian for modified gravity $f(T,\mathcal{B})$ is defined as \cite{Abedi2017}
\begin{eqnarray}
\mathcal{L}= \frac{h}{4\kappa_g}f(T,\mathcal{B}).
\end{eqnarray}
Thus, the assumed gravitational action in our six-dimensional model is
\begin{eqnarray}\label{55.5}
\mathcal{S}=-\frac{1}{4\kappa_g}\int h f(T,\mathcal{B})d^6x+\int h\left(\Lambda +\mathcal{L}_m\right)d^6x,
\end{eqnarray}
where $\mathcal{L}_m$ is the matter Lagrangian.

The action (\ref{55.5}) leads us to the following field equations \cite{Abedi2017,Pourbagher2020}
\begin{eqnarray}\label{3.36}
\frac{1}{h}f_T\Big[\partial_Q(h S_N\ ^{MQ})-h\widetilde{\Gamma}^R\ _{SN}S_R\ ^{MS}\Big]+\frac{1}{4}\Big[f-\mathcal{B}f_\mathcal{B}\Big]\delta_N^M& &\nonumber\\+\Big[(\partial_Qf_T)+(\partial_Qf_\mathcal{B}) \Big]S_N\ ^{MQ} +\frac{1}{2}\Big[\nabla^M\nabla_N f_\mathcal{B}-\delta^M_N\Box f_\mathcal{B}\Big]&=&-\kappa_g(\Lambda\delta_N^M+\mathcal{T}_N\ ^M),\nonumber\\
\end{eqnarray}
where $\Box\equiv\nabla^M\nabla_M$, $f\equiv f(T,\mathcal{B})$, $f_T\equiv\partial f(T,\mathcal{B})/\partial T$ e $f_{\mathcal{B}}\equiv\partial f(T,\mathcal{B})/\partial \mathcal{B}$.

The metric of a string-like braneworld senary can be described by the ansatz \cite{Gherghetta,Liu,Moreira:2020rni,Moreira:2021fva}
\begin{equation}\label{45.a}
ds^2=e^{2A(r)}\eta_{\mu\nu}dx^\mu dx^\nu+dr^2+R^2_0e^{2B(r)}d\theta^2,
\end{equation}
where $0 \leq r \leq r_{max} $, $\theta \in [0; 2\pi)$, 
being $\eta_{\mu\nu}=(-1, 1, 1, 1)$ is the well-known metric of
Minkowski that describes the space we live in. Here $e^{A(r)}$ and $e^{B(r)}$ are called warp factors that satisfy the conditions
\begin{eqnarray}
\label{regularityconditions}
e^{A(0)}=1 &,& (e^{A})'(0)=0,\nonumber\\
e^{B(0)}=0 &,& (e^{B})'(0)=1,
\end{eqnarray}
ensuring smooth geometry at the origin \cite{Giovannini:2001hh,Liu}.

For the metric (\ref{45.a}) we select the \textit{sechsbeins}
$h_a\ ^M=diag(e^A, e^A, e^A, e^A, 1, R_0 e^B)$, which represents a good choice among all vielbeins \cite{Ferraro2011us, Tamanini2012}. Thus, the torsion scalar and the boundary term take the form
\begin{eqnarray}
 T&=&-4A'(3A'+2B'),\nonumber\\
 \mathcal{B}&=&-2[(4A'+B')^2+4A''+B''],
\end{eqnarray}
where the prime $(\ '\ )$ denotes differentiation with respect to $r$.

The term $\mathcal{T}_N\ ^M$ in equation (\ref{3.36}) is the stress-energy tensor. Here, due to the axisymmetric brane geometry we assume a stress-energy tensor in the form
\begin{equation}
\mathcal{T}_{MN}dx^M \otimes dx^N=t_0(r)dx^\mu\otimes dx^\nu + t_r (r) dr\otimes dr + t_\theta (r) d\theta\otimes d\theta, 
\end{equation}
{where the component $t_0$ represents the energy density of the brane, $t_r$ the radial pressure and $t_\theta$ the angular pressure.}

In this way, the equations of motion become
\begin{eqnarray}\label{e.1}
\frac{1}{2}\Big\{\Big[(4A'+B')^2+4A''+B''\Big]f_\mathcal{B}+(3A'+B')(f_\mathcal{B}'+f_T')\Big\}&&\nonumber\\ +\frac{1}{2}\Big[(4A'+B')(3A'+B')+3A''+B''\Big]f_{T}+\frac{1}{4}f&=&-(\Lambda+t_0),\nonumber\\ \\  
\label{e.2}
\frac{1}{2}\Big[(4A'+B')^2+4A''+B''\Big]f_\mathcal{B}+2A'(f_\mathcal{B}'+f_T')&&\nonumber\\ +2\Big[A'(4A'+B')+A''\Big]f_{T}+\frac{1}{4}f&=&-( \Lambda+t_\theta),\nonumber\\ \\  
\label{e.3}
\frac{1}{2}\Big[(4A'+B')^2+4A''+B''\Big]f_\mathcal{B}+2A'\Big(3A'+2B'\Big)f_T+\frac{1}{4}f&=&-(\Lambda+t_r),\nonumber\\ 
\end{eqnarray}
where $\kappa_g=1$ for simplicity.

With the equations of motion (\ref{e.1}, \ref{e.2} and \ref{e.3}) it remains for us to define the form of the modified gravities $f(T,\mathcal{B}) $. As we aim to study generalized forms of conventional teleparallelism, we selected three specific descriptions of modified gravities, i. e., $f_1(T,\mathcal{B})=T+k\mathcal{B}^{n}$, $f_2(T,\mathcal{B})=T+k(-T+\mathcal{B})^{n}$ and $f_3(T,\mathcal{B})=T+k_1T^2+k_2\mathcal{B}^2$, where $n$, $k$ and $k_{1,2}$ are parameters that control the deviation from the usual teleparallelism. Note that if we did $k=k_1=k_2=0$ we fall back on TEGR.

{
As a matter of fact, we selected the gravity models $f_{1,2,3}$ based on the significant results already presented in the literature. The $f_{1,3}$ models have worked well in studying the accelerated expansion of the Universe, cosmological perturbations, and thermodynamics \cite{Wright2016,Bahamonde2016,Franco2020,EscamillaRivera2019,Bahamonde2016a}. The $f_2$ model, on the other hand, presented relevant results in a Noether symmetry approach in the cosmological scenario \cite{Bahamonde2016}.}

{
It is noteworthy that neither the torsion scalar $T$ nor the boundary term $B$ are invariant under local Lorentz transformations. This can be useful when studying metric theories of gravity that already break this invariance \cite{Bahamonde2015,Jacobson:2007veq}. Furthermore, for a nonlinear function $f(T)$, gravity $f(T)$ is the only possible second-order theory of modified gravity, but the price to pay is the violation of local Lorentz invariance. This is not a deficit, and the theory, although not the Lorentz covariant, is significant. }

{
In a sense, the limit term $B$ can be used to combine with $T$ to form a covariant Lorentz theory. Note that if we set $k=-1$ and $n=1$ in $f_1(T,B)$, we fall back into a theory $f(R)$ which is manifestly Lorentz invariant. Therefore, we can conclude that the field equations are Lorentz invariants if and only if they are equivalent to gravity $f(R)$. So the teleparallel equivalent of gravity $f(R)$ is the only possible invariant Lorentz theory of gravity constructed from $R$, $T$, and $B$.
}

For smooth thick string-like solutions, we propose a warp factor ansatz in the form
\begin{eqnarray}
\label{coreA}
A(r)&=&\ln[\mathrm{sech}^{p}(\lambda r)],\nonumber\\
\label{coreB}
B(r)&=&\ln[\mathrm{sech}^{p}(\lambda r)]+\ln[\tanh(\rho r)],
\end{eqnarray}
where the regularity conditions (\ref{regularityconditions}) are satisfied. The parameters $p$, $\lambda$ and $\rho$ determine the breadth and width of the source.

We obtain the value of the cosmological constant knowing that asymptotically ($r\rightarrow\infty$) the identity $\mathcal{T}_N\ ^M= 0$ is satisfied. So, for $f_1$ we get
\begin{eqnarray}
\Lambda=-\frac{1}{4}\left[20p^2\lambda^2+(-1)^{n+1}k(n-1)\left(50p^2\lambda^2\right)^{n}\right].
\end{eqnarray}
For $f_2$ we have
\begin{eqnarray}
\Lambda=-\frac{1}{12}\left[60p^2\lambda^2+(-1)^{n+1}k(n-3)\left(30p^2\lambda^2\right)^{n}\right].
\end{eqnarray}
Finally, for $f_3$
\begin{eqnarray}
\Lambda=-5p^2\lambda^2\left[1+5(12k_1+25k_2)p^2\lambda^2\right].
\end{eqnarray}
 
Furthermore, we can analyze the behavior of energy densities and pressures for each function $f(T,\mathcal{B})$ established. Eqs.(\ref{e.1}, \ref{e.2} and \ref{e.3}) take a large form for each model $f_{1,2,3}$. That is why we decided to demonstrate only the respective plots of the graphic behavior.

\begin{figure}[ht!]
\begin{center}
\centering
\begin{tabular}{ccc}
\includegraphics[height=4cm]{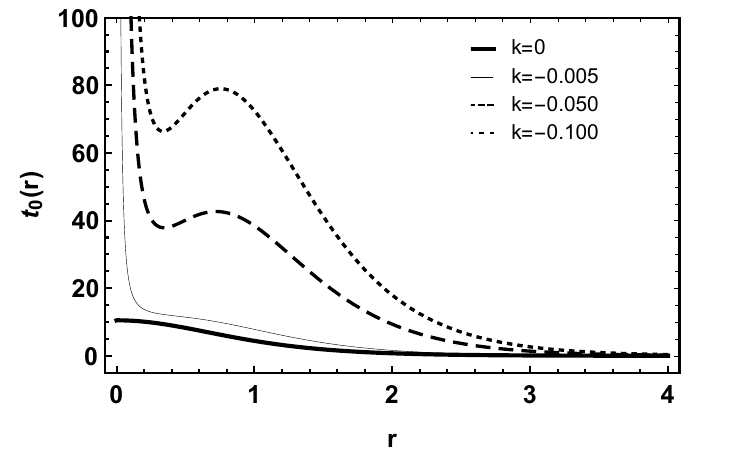} 
\includegraphics[height=4cm]{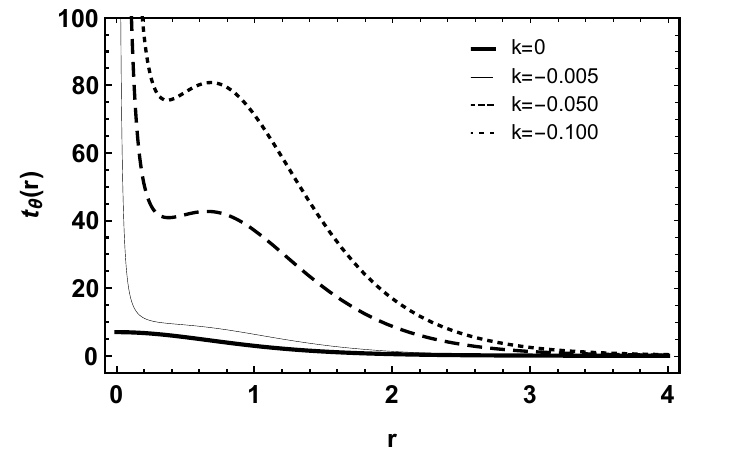}\\
(a) \hspace{8 cm}(b)\\
\includegraphics[height=4cm]{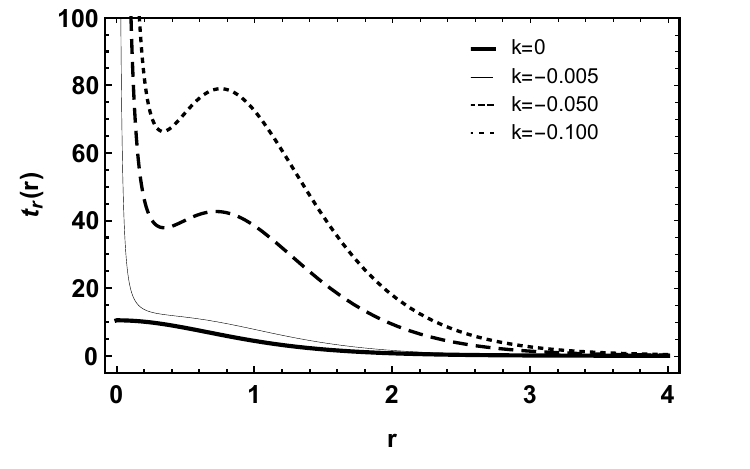}\\
(c) 
\end{tabular}
\end{center}
\caption{For $f_1$ where $n=2$ with $\lambda=p=\rho=1$. (a) Energy density. (b) Angular pressure. (c) Radial pressure.
\label{fig1}}
\end{figure}

In Fig.\ref{fig1} we plot the graphical behavior of the energy density and the respective angular and radial pressures for $f_1$. It is interesting how the boundary term modifies the energy density. When we consider $k=0$ (without the contribution of $\mathcal{B}$) we obtain the TERG result, where the density has a single peak centered at $r=0$. But when $k\neq0$, the peak at the core turns out to be a super-localized tower. Also, when we make $k$ more negative, a new peak forms right after the super-localized tower. The same goes for pressures.

\begin{figure}[ht!]
\begin{center}
\begin{tabular}{ccc}
\includegraphics[height=4cm]{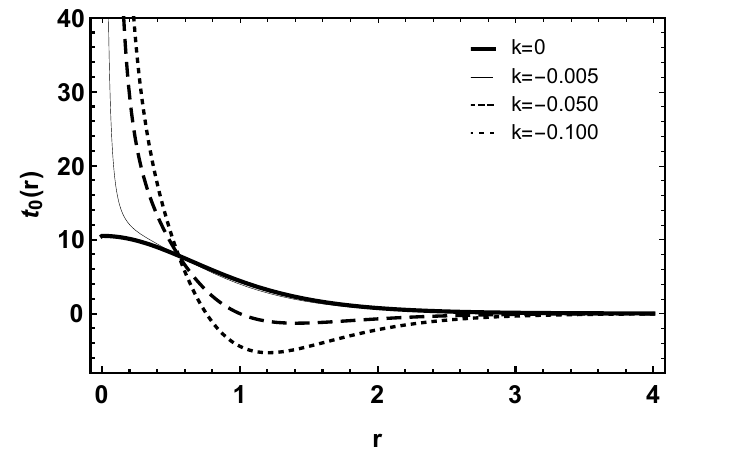} 
\includegraphics[height=4cm]{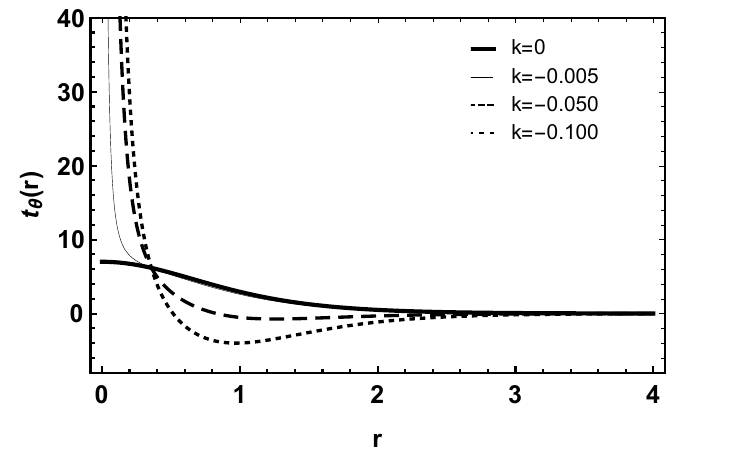}\\
(a) \hspace{8 cm}(b)\\
\includegraphics[height=4cm]{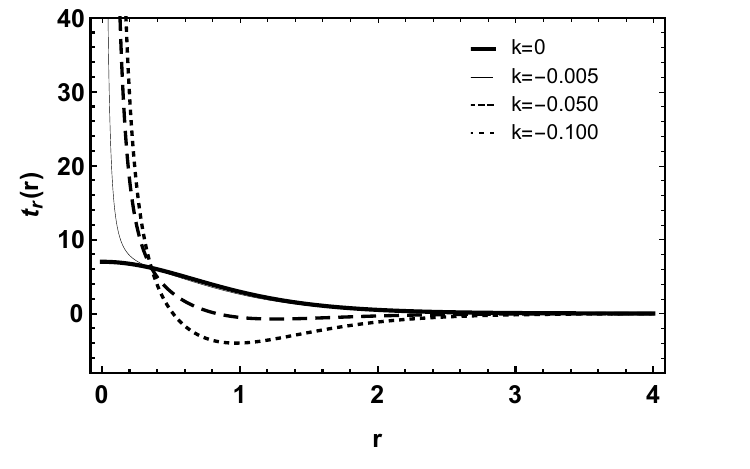}\\
(c) 
\end{tabular}
\end{center}
\caption{For $f_2$ where $n=2$ with $\lambda=p=\rho=1$. (a) Energy density. (b) Angular pressure. (c) Radial pressure.
\label{fig2}}
\end{figure}

The choice of the function $f_2(T,\mathcal{B})=T+k(-T+\mathcal{B})^{n}$ is interesting because it can be understood as a function of $f(T,R )=T+kR^n$, remembering that $R=-T+\mathcal{B}$. Thus, in Fig.\ref{fig2} we plot the behavior of the energy density and the respective angular and radial pressures. When we consider the influence of $R$ on the energy density, we observe that the peak centered on the core becomes a super-localized tower. Furthermore, the more negative $k$ is, we observe the appearance of a negative phase in the density. The same is repeated in pressures.

\begin{figure}[ht!]
\begin{center}
\begin{tabular}{ccc}
\includegraphics[height=4cm]{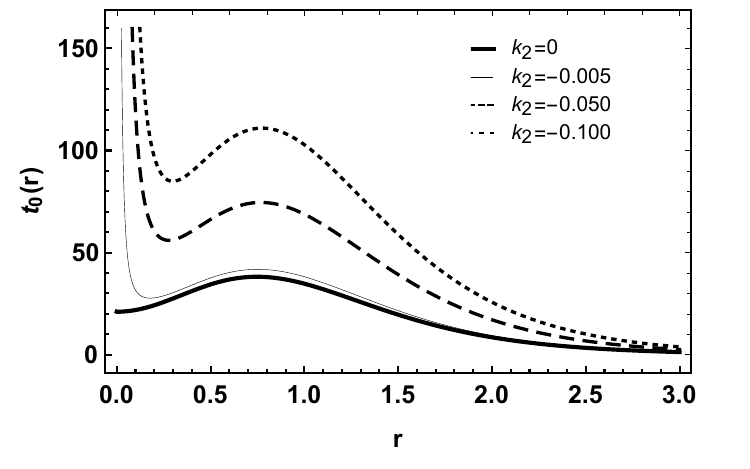} 
\includegraphics[height=4cm]{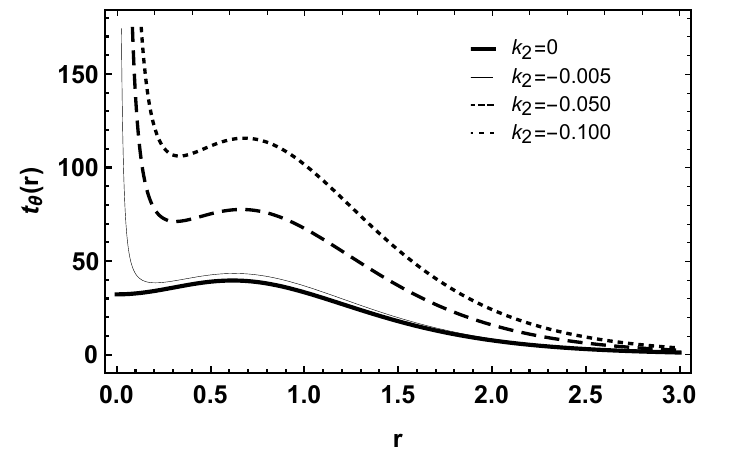}\\
(a) \hspace{8 cm}(b)\\
\includegraphics[height=4cm]{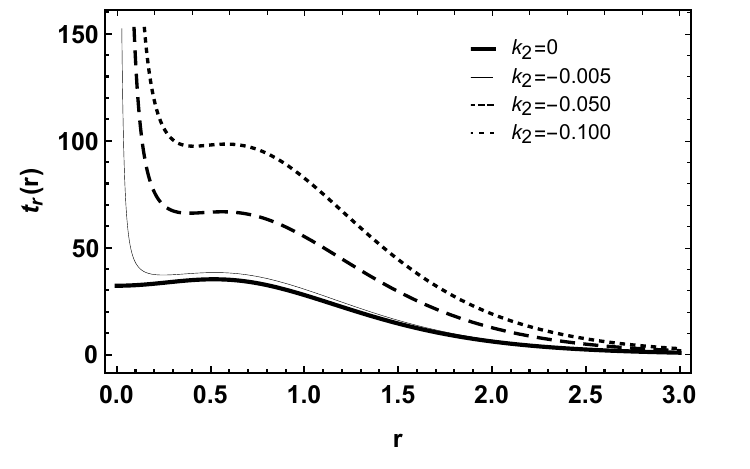}\\
(c) 
\end{tabular}
\end{center}
\caption{For $f_3$ where $k_1=-0.1$ with $\lambda=p=\rho=1$. (a) Energy density. (b) Angular pressure. (c) Radial pressure.
\label{fig3}}
\end{figure}

In function $f_3$ we can analyze the influence of torsion and boundary term on energy density and angular and radial pressures. In Fig.\ref{fig3} we analyze the joint influence of torsion ($k_1=-0.1$) and the boundary term. Note that again when we consider the boundary term, we get the appearance of a super-localized tower in the core. Furthermore, the smaller the $k_2$, the more evident the ring-like structure becomes after the tower is located.

\begin{figure}[ht!]
\begin{center}
\begin{tabular}{ccc}
\includegraphics[height=4cm]{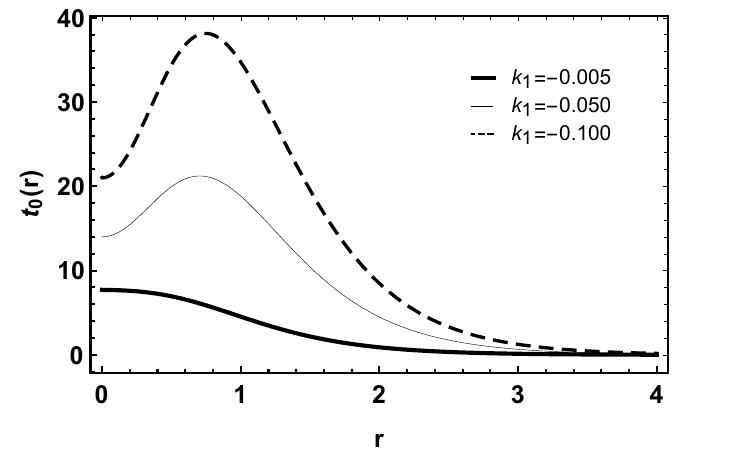} 
\includegraphics[height=4cm]{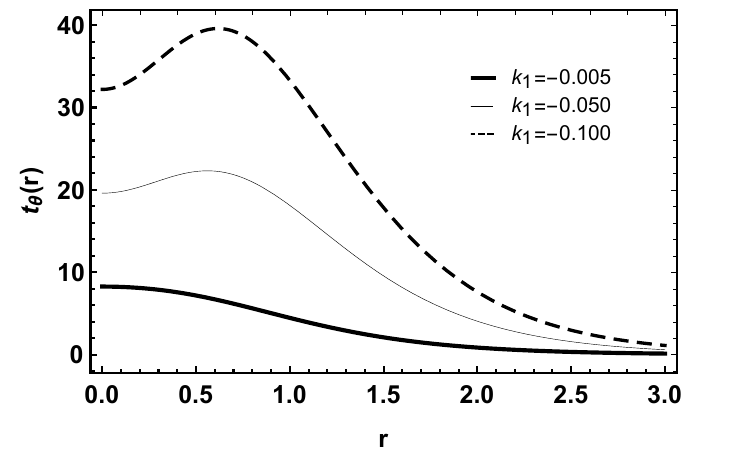}\\
(a) \hspace{8 cm}(b)\\
\includegraphics[height=4cm]{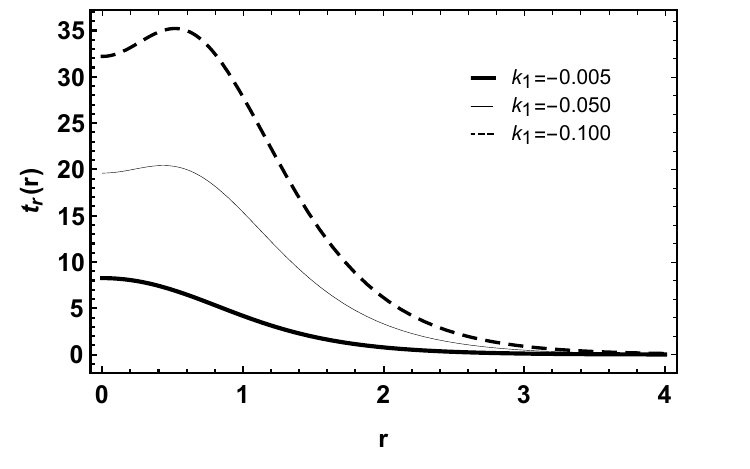}\\
(c) 
\end{tabular}
\end{center}
\caption{For $f_3$ where $k_2=0$ with $\lambda=p=\rho=1$. (a) Energy density. (b) Angular pressure. (c) Radial pressure.
\label{fig4}}
\end{figure}

When we disregard the influence of the boundary term ($k_2=0$), we can analyze the influence of torsion on the energy density. From Fig.\ref{fig4} it is evident that when we consider more negative values of $k_1$, a ring-like structure is formed right after the core. This result is repeated in pressures.

{When we vary the parameter that controls the influence of the boundary term, new maxima appear in the energy density plot (Figs.\ref{fig1}, \ref{fig2} and \ref{fig3}). These new maxima equate to the emergence of ring-like structures that are most evident in functions $f_1$ and $f_3$. The appearance of the ring-like structure in the energy density indicates the brane split. Furthermore, the term that controls the influence of torsion on the $f_3$ function can reproduce the appearance of the ring-like structure (see \cite{Moreira:2020rni,Moreira:2021fva}).
}

\section{Gravitational Perturbations and localization}
\label{sec2}

In this section, we investigate the effects of varying the boundary term and the torsion scalar on the propagation of gravitational perturbations. We then consider the \textit{seschsbein} perturbation \cite{Moreira:2020rni,Moreira:2021fva}
\begin{eqnarray}\label{000588}
h^a\ _M=\left(\begin{array}{cccccc}
e^{A(r)}\left(\delta^a_\mu+w^a\ _\mu\right)&0&0\\
0&1&0\\
0&0&R_0e^{B(r)}\\
\end{array}\right),
\end{eqnarray}
which leads us to the resulting metric perturbation
\begin{eqnarray}
ds^2=e^{A(r)}\left(\eta_{\mu\nu}+\gamma_{\mu\nu}\right)dx^\mu dx^\nu+dr^2+R^2_0e^{2B(r)}d\theta^2
\end{eqnarray}
where $w^a\ _\mu=w^a\ _\mu(x^\mu,r,\theta)$ 
and the relation 
\begin{eqnarray}
\gamma_{\mu\nu}=(\delta^a_\mu w^b\ _\nu+\delta^b_\nu w^a\  _\mu)\eta_{ab},
\end{eqnarray}
is obeyed. Here, we assume the transverse-traceless (TT) metric gauge satisfies the following condition
\begin{eqnarray}
\partial_\mu \gamma^{\mu\nu}=0=\eta^{\mu\nu}\gamma_{\mu\nu},
\end{eqnarray}
which guarantees a \textit{sechsbein} gauge $\delta_a^\mu w^a\  ^\mu=0$ \cite{Moreira:2020rni,Moreira:2021fva}. These conditions lead us to $\delta h=0$, $\delta T=0$ and $\mathcal{B}=0$.

The form of the equation (\ref{3.36}) perturbed is
\begin{eqnarray}\label{27.l}
\frac{1}{4}(\mathcal{B}f_\mathcal{B}-f)\delta g_{MN}-\frac{1}{h}f_T\Big[\delta g_{PN}\partial_Q(h S_M\ ^{PQ})+ g_{PN}\partial_Q(h \delta S_M\ ^{PQ})&\nonumber\\
-h\Big(\delta\widetilde{\Gamma}^Q\ _{PM}S_{QN}\ ^{P}+\widetilde{\Gamma}^Q\ _{PM}\delta S_{QN}\ ^{P}\Big)\Big] -\Big[(\partial_Qf_\mathcal{B})+(\partial_Qf_T)\Big]\delta S_{MN}\ ^{Q}\nonumber\\
+\frac{1}{2}\Big(\delta g_{MN}\Box f_\mathcal{B}-\delta g_{PN}\nabla^P\nabla_Mf_\mathcal{B}\Big)&=&\delta\mathcal{T}_{MN}.\nonumber\\
\end{eqnarray}
It is important to note that the perturbations disappear in the extra dimensions, leaving only the brane perturbation
\begin{eqnarray}\label{28.l}
\delta\mathcal{T}_{\mu\nu}&=&\frac{e^{2A}}{4}\Big\{e^{-2A}\Box^{(4)} \gamma_{\mu\nu}+(4A'+B')\gamma'_{\mu\nu}+\gamma''_{\mu\nu}-2\Big[(4A'+B')(3A'+B')\nonumber\\ &+&3A''+B''\Big]\gamma_{\mu\nu}+R_0^{-2}e^{-2B}\partial_\theta^2\gamma_{\mu\nu}\Big\}f_T
-\frac{e^{2A}}{4}\Big\{f+2\Big[(4A'+B')^2\nonumber\\
&+&4A''+B''\Big]f_\mathcal{B}\Big\}\gamma_{\mu\nu}+e^{2A}\Big[\frac{1}{2}(3A'+B')\gamma_{\mu\nu}-\frac{1}{4}\gamma'_{\mu\nu}\Big]\Big\{8\Big[A'(3A''+B'')\nonumber\\
&+&A''B'\Big](f_{TT}+f_{\mathcal{B}T})+4\Big[(4A'+B')(4A''+B'')\nonumber\\
&+&2A'''+\frac{1}{2}B'''\Big](f_{\mathcal{B}\mathcal{B}}+f_{T\mathcal{B}})\Big\},
\end{eqnarray}
where $\Box^{(4)}=\eta^{\mu\nu}\partial_\mu \partial_\nu$. The perturbation of the stress-energy tensor is
\begin{eqnarray}\label{29.l}
\delta\mathcal{T}_{\mu\nu}=\delta(\mathcal{T}_{\mu}\ ^\mu g_{\mu\nu})=\delta(\mathcal{T}_{\mu}\ ^\mu)\eta_{\mu\nu}e^{2A}+\mathcal{T}_{\mu}\ ^\mu \gamma_{\mu\nu}e^{2A},
\end{eqnarray}
where the trace $\delta(\mathcal{T}_{\mu}\ ^\mu)$ vanish. 
From Eq.(\ref{3.36}) we get that
\begin{eqnarray}\label{30.l}
\mathcal{T}_{\mu}\  ^\mu&=&
\frac{1}{2}(3A'+B')\Big\{4\Big[(4A'+B')(4A''+B'')+2A'''+\frac{1}{2}B'''\Big](f_{\mathcal{B}\mathcal{B}}+f_{T\mathcal{B}})\nonumber\\
&+&\Big[24A''A'+8(A'B''+A''B')\Big](f_{TT}+f_{\mathcal{B}T})\Big\}-\frac{1}{4}\Big\{f+2\Big[(4A'+B')^2\nonumber\\
&+&4A''+B''\Big]f_\mathcal{B}\Big\}-\frac{1}{2}\Big[(4A'+B')(3A'+B')+3A''+B''\Big]f_{T}.
\end{eqnarray}

The Eqs.(\ref{29.l}) and (\ref{30.l}) give the perturbation equation
\begin{eqnarray}\label{32.l}
[e^{-2A}\Box^{(4)} \gamma_{\mu\nu}+(4A'+B')\gamma'_{\mu\nu}+\gamma''_{\mu\nu}+R_0^{-2}e^{-2B}\partial_\theta^2\gamma_{\mu\nu}]f_T& &\nonumber\\ -4\Big\{\Big[(4A'+B')(4A''+B'')+2A'''+\frac{1}{2}B'''\Big](f_{\mathcal{B}\mathcal{B}}+f_{T\mathcal{B}})& &\nonumber\\ +\Big[6A''A'+2(A'B''+A''B')\Big](f_{TT}+f_{\mathcal{B}T})\Big\}\gamma'_{\mu\nu}&=&0.
\end{eqnarray}
By making the conformal coordinate transformation
\begin{eqnarray}
dz=e^{-A}dr,
\end{eqnarray}
Eq.(\ref{32.l}) takes the form
\begin{eqnarray}\label{32.66}
\frac{f_T}{4}\Big[\Box^{(4)}+\partial^2_z+R_0^{-2}e^{-2B}\partial_\theta^2+(3\dot{A}+\dot{B})\Big]\gamma_{\mu\nu}+e^{-2A}\Big\{\Big[\frac{1}{2}\Big(\dddot{B}& &\nonumber\\ -3\dot{A}\ddot{B}-\ddot{A}\dot{B}+\dot{B}\dot{A}^2\Big)+2(\dddot{A}-4\dot{A}\ddot{A}+2\dot{A}^3)+(4\dot{A}+\dot{B})\Big(4(\ddot{A}-\dot{A}^2)& &\nonumber\\ +3(\ddot{B}-\dot{A}\dot{B})\Big)\Big](f_{\mathcal{B}\mathcal{B}}+f_{T\mathcal{B}}) +
\Big[(\dot{A}+\dot{B})(\ddot{A}-\dot{A}^2)& &\nonumber\\ +\dot{A}(\ddot{B}-\dot{A}\dot{B})\Big](f_{TT}+f_{\mathcal{B}T})\Big\}\partial_z\gamma_{\mu\nu}&=&0,\nonumber\\ 
\end{eqnarray}
where the dot $(\ \dot{}\ )$  denotes differentiation with respect to $z$.

We can rewrite Eq.(\ref{32.66}) as
\begin{eqnarray}
(\Box^{(4)}+2H\partial_z+\partial^2_z+R_0^{-2}e^{-2B}\partial_\theta^2)\gamma_{\mu\nu}=0,
\end{eqnarray}
where 
\begin{eqnarray}\label{32.56}
H(z)&=&\frac{1}{2}(3\dot{A}+\dot{B})+\frac{2}{f_T}e^{-2A}\Big\{\Big[\frac{1}{2}\Big(\dddot{B} -3\dot{A}\ddot{B}-\ddot{A}\dot{B}+\dot{B}\dot{A}^2\Big)\nonumber\\&+&2(\dddot{A}-4\dot{A}\ddot{A}+2\dot{A}^3)+(4\dot{A}+\dot{B})\Big(4(\ddot{A}-\dot{A}^2)\nonumber\\ &+&3(\ddot{B}-\dot{A}\dot{B})\Big)\Big](f_{\mathcal{B}\mathcal{B}}+f_{T\mathcal{B}}) +
\Big[(\dot{A}+\dot{B})(\ddot{A}-\dot{A}^2)\nonumber\\ &+&\dot{A}(\ddot{B}-\dot{A}\dot{B})\Big](f_{TT}+f_{\mathcal{B}T})\Big\}.
\end{eqnarray}

We can now introduce the KK decomposition 
\begin{equation}
\gamma_{\mu\nu}(x^\rho,r,\theta)=\epsilon_{\mu\nu}(x^\rho)\sum_{\beta=0}^{\infty} \chi(z)\Psi(z) e^{i\beta\theta}, 
\end{equation}
where $\chi(z)=e^{-\frac{1}{2}(3A+B)+\int K(z)dz}$ with
\begin{eqnarray}
K(z)&=&-\frac{2}{f_T}e^{-2A}\Big\{\Big[\frac{1}{2}\Big(\dddot{B} -3\dot{A}\ddot{B}-\ddot{A}\dot{B}+\dot{B}\dot{A}^2\Big)\nonumber\\&+&2(\dddot{A}-4\dot{A}\ddot{A}+2\dot{A}^3)+(4\dot{A}+\dot{B})\Big(4(\ddot{A}-\dot{A}^2)\nonumber\\ &+&3(\ddot{B}-\dot{A}\dot{B})\Big)\Big](f_{\mathcal{B}\mathcal{B}}+f_{T\mathcal{B}}) +
\Big[(\dot{A}+\dot{B})(\ddot{A}-\dot{A}^2)\nonumber\\ &+&\dot{A}(\ddot{B}-\dot{A}\dot{B})\Big](f_{TT}+f_{\mathcal{B}T})\Big\}.
\end{eqnarray}
Also, the 4D plane-wave condition $\left(\Box^{(4)}-m^2\right)\epsilon_{\mu\nu}=0$ is satisfied. All this leads us to a simplified Sch\"{o}dinger-like equation 
\begin{eqnarray}\label{36.l}
\left[-\partial_z^2+U(z)\right]\Psi(z)=m^2\Psi(z),
\end{eqnarray}
where the potential is defined by $U(z)=\dot{H}+H^2+\beta^2R_0^{-2}e^{2(A-B)}$.

Setting $\beta=0$, the $U$ potential falls into the supersymmetric form of quantum mechanics. This choice promotes the absence of tachyonic KK gravitational modes, ensuring the stability of the spectrum, in addition to presenting well-localized massless modes in the form
\begin{eqnarray}
\Psi_0=N_0e^{\frac{1}{2}(3A+B))-\int K(z)dz},
\end{eqnarray}
where $N_0$ is a normalization constant. 

The effective potential forms for functions $f_{1,2,3}$ are too big to write here. Therefore, we analyzed the graphical behavior of these potentials. Furthermore, we analyze the massless modes.

\begin{figure}[ht!]
\begin{center}
\begin{tabular}{ccc}
\includegraphics[height=4cm]{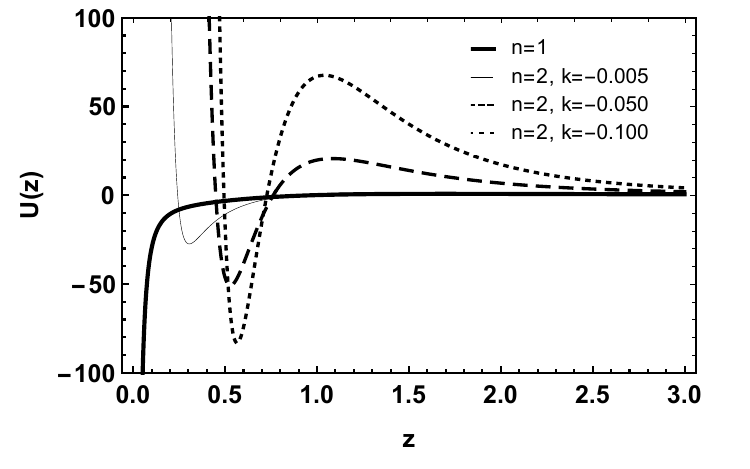} 
\includegraphics[height=4cm]{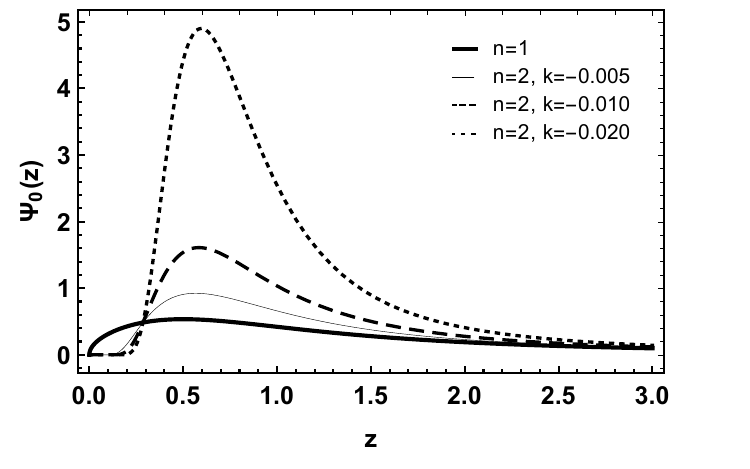}\\
(a) \hspace{8 cm}(b)
\end{tabular}
\end{center}
\caption{For $f_1$ with $\lambda=p=\rho=1$. (a) Effective potential. (b) Massless mode. 
\label{fig5}}
\end{figure}

For $f_1$ with $n=1$ the effective potential becomes invariant under the parameter $k$. This happens because the term $K=0$ disappears when $n=1$, which takes us to the TEGR case, where the potential is an infinite well at the origin. For $n=2$, the term $K\neq0$, makes it possible to analyze the influence of the boundary term on the effective potential, which becomes an infinite potential barrier at the origin. When we decrease the value of $k$, we notice the emergence of new wells and finite potential barriers, right after the infinite barrier. Massless modes sense this change in effective potential. When we vary the value of $k$, we make the massless modes more localized with a peak that moves away from the origin, forming ring-like structures (See Fig.\ref{fig5}).

Something interesting happens for $f_2$. The term $K$ disappears for any value of $k$ and $n$, falling back to TEGR. The graphic behavior is the same as the one presented for $f_1$ with $n=1$ (Fig.\ref{fig5}).

\begin{figure}[ht!]
\begin{center}
\begin{tabular}{ccc}
\includegraphics[height=4cm]{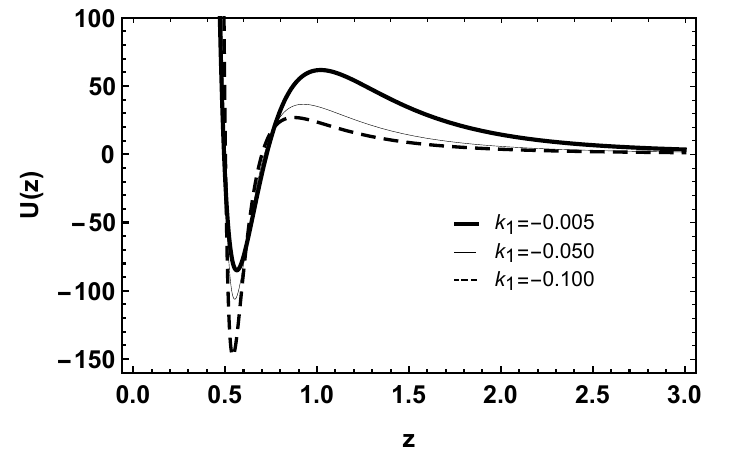} 
\includegraphics[height=4cm]{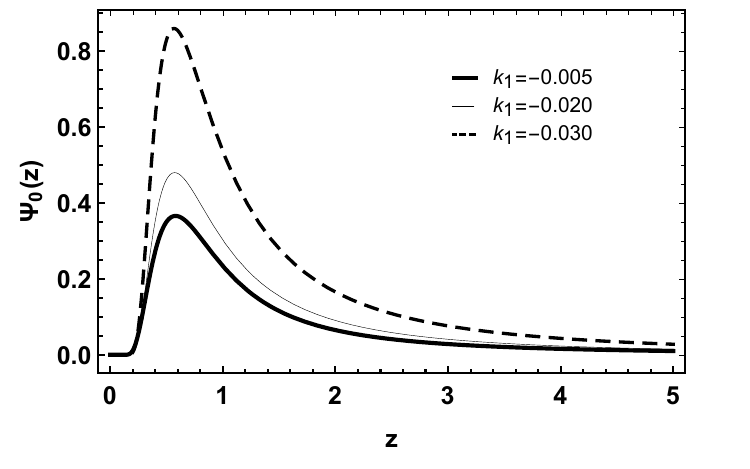}\\
(a) \hspace{8 cm}(b)
\end{tabular}
\end{center}
\caption{For $f_3$ where $k_2=-0.1$ with $\lambda=p=\rho=1$. (a) Effective potential. (b) Massless mode. 
\label{fig6}}
\end{figure}

\begin{figure}[ht!]
\begin{center}
\begin{tabular}{ccc}
\includegraphics[height=4cm]{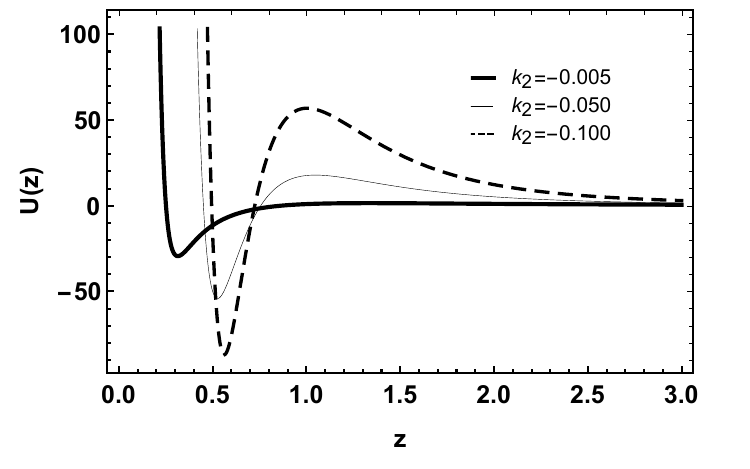} 
\includegraphics[height=4cm]{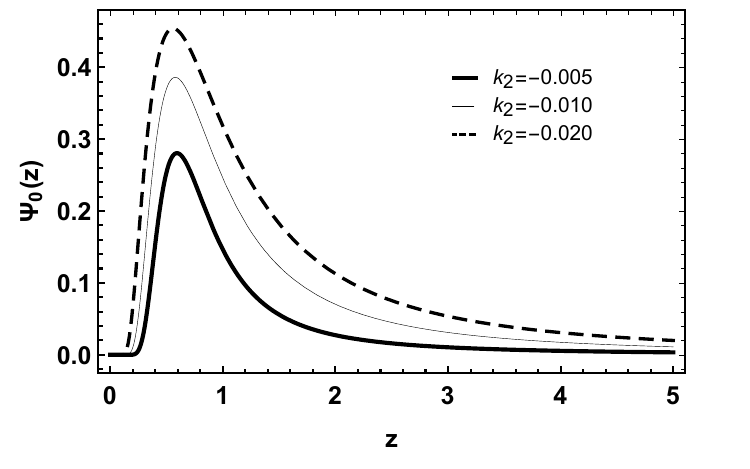}\\
(a) \hspace{8 cm}(b)
\end{tabular}
\end{center}
\caption{For $f_3$ where $k_1=-0.01$ with $\lambda=p=\rho=1$. (a) Effective potential. (b) Massless mode. 
\label{fig7}}
\end{figure}

As for $f_3$, we can analyze the influence of torsion and the boundary term. In Fig.\ref{fig6} it is possible to observe that when we decrease the value of the torsion parameter $k_1$ we intensify the finite potential well, directly affecting the behavior of the massless modes that become more localized. In Fig.\ref{fig7} it is possible to notice the influence of the boundary parameter $k_2$, which in addition to intensifying the well and finite potential barrier, tends to shift them away from the origin. The massless modes sense the effective potential change.

\begin{figure}[ht!]
\begin{center}
\begin{tabular}{ccc}
\includegraphics[height=4cm]{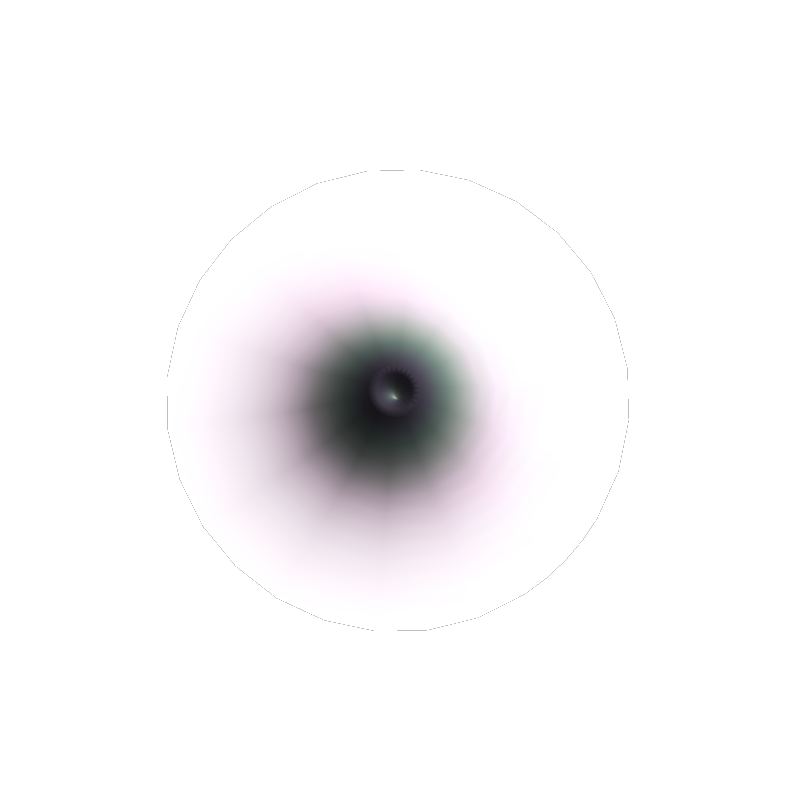} 
\includegraphics[height=4cm]{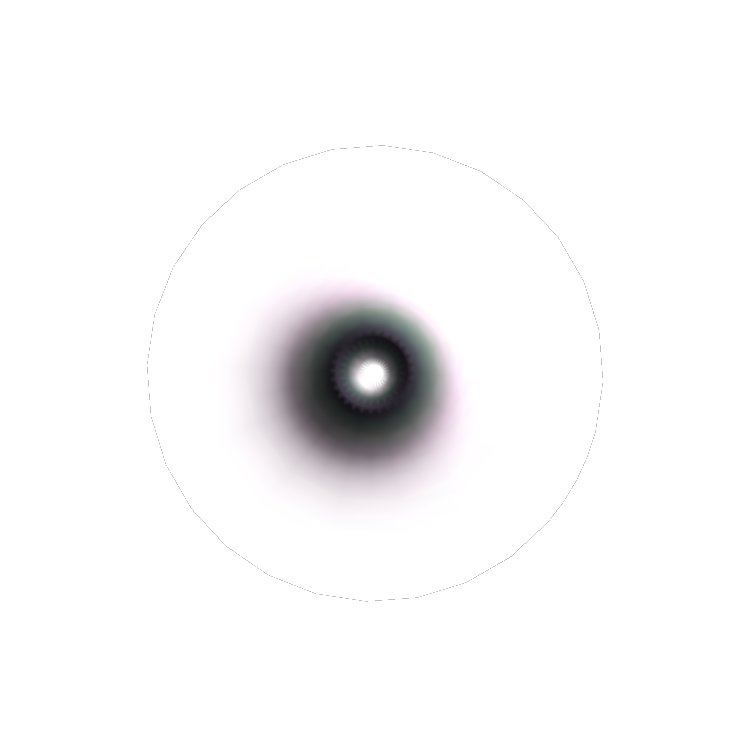}
\includegraphics[height=4cm]{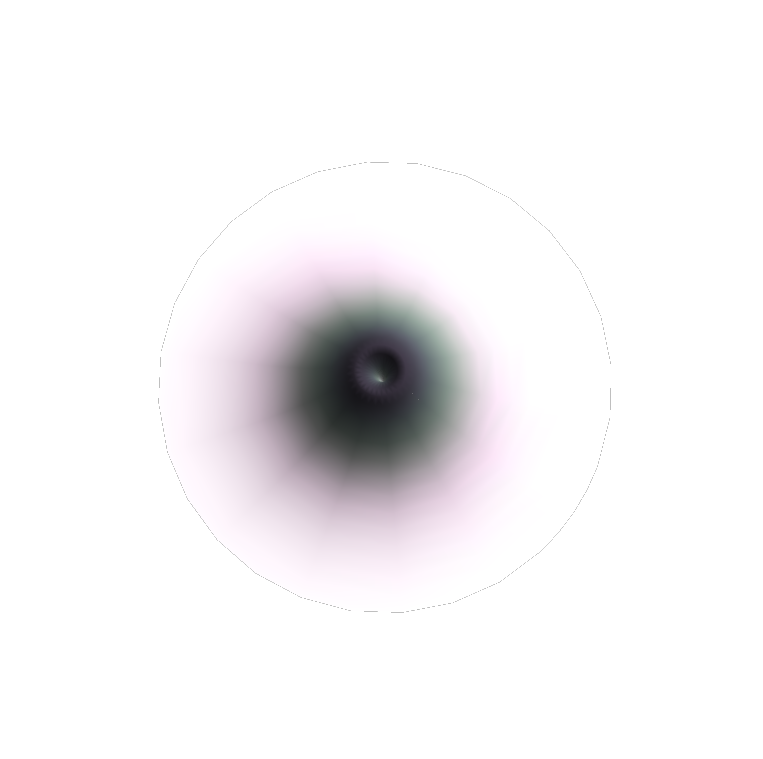}\\
(a) \hspace{3.5 cm}(b) \hspace{3.5 cm}(c)
\end{tabular}
\end{center}
\caption{Massless modes in the $z-\theta$ plane, where $\lambda=p=\rho=1$. (a) $f_1$ with $n=1$. (b) $f_1$ with $n=2$ and $k=-0.02$. (c) $f_3$ with $k_1=-0.01$ and $k_2=-0.02$. 
\label{fig9}}
\end{figure}

{We can see the behavior of the massless mode in the plane $z-\theta$ just by revolutionizing the plot of $\Psi_0$ around the coordinate $\theta$. See figure \ref{fig9}.}

\section{Conclusion}
\label{finalremarks} 
 
In this work, we study the effects of the boundary term $\mathcal{B}$ on a string-like braneworld in the context of the $f(T,\mathcal{B})$ gravity.

Through the energy densities and the respective angular and radial pressures, it was possible to observe the influence of the boundary term that tends to form a super-localized structure in the core. Furthermore, new maximum and minimum points appear, leading to negative pressures, thus violating the dominant energy and weak energy conditions. 
{These new peaks in energy density form a ring-like structure that indicates the brane split \cite{Moreira:2020rni,Moreira:2021fva}.}
Therefore the boundary term can produce modifications in the source equation of state that represent the brane split. Furthermore, for $f_3$ in particular, the influence of torsion was analyzed, which can produce a ring-like structure, also symbolizing the splitting of the brane.

It was possible to analyze the influence of the boundary term on gravitational perturbations. We arrive at an equation similar to that of Sch\"{o}dinger, where the effective potential presents the supersymmetric form of quantum mechanics for $\beta=0$. This guarantees the absence of tachyonic KK gravitational modes which leads to the stability of the spectrum. The boundary term transforms the infinite well of effective potential into an infinite barrier of potential, in addition to introducing new wells and finite barriers right after the infinite barrier. The massless modes become more localized as we intensify the influence of the boundary term. The same happens when we analyze the influence of torsion.

These results aroused our curiosity about the influence of the boundary term on the behavior of massive modes. Furthermore, a future perspective would be to analyze the possibility of the existence of resonant modes.

\section*{Acknowledgments}
\hspace{0.5cm} The authors thank the Conselho Nacional de Desenvolvimento Cient\'{\i}fico e Tecnol\'{o}gico (CNPq), grants n$\textsuperscript{\underline{\scriptsize o}}$ 309553/2021-0 (CASA), for financial support. {The authors also thank the anonymous referee for their valuable comments and suggestions.}


\subsection*{Conflicts of interest/Competing interest}
All the authors declared that there is no conflict of interest in this manuscript.

\section*{Data Availability}
The datasets generated during and/or analysed during the current study are available from the corresponding author on reasonable request


\end{document}